\documentclass[conference,compsoc]{IEEEtran}
\ifCLASSOPTIONcompsoc
\usepackage[nocompress]{cite}
\else
\usepackage{cite}
\fi
\usepackage{verbatim}
\usepackage{enumitem}
\usepackage{hyperref}
\usepackage{graphicx}
\usepackage{subcaption}
\usepackage{booktabs}
\usepackage{multirow}
\usepackage{xspace}
\usepackage{array}
\usepackage{xcolor}
\usepackage{tabularx}
\usepackage{makecell}
\usepackage{float}
\usepackage{colortbl}  
\usepackage{xcolor}    
\usepackage{textcomp}
\usepackage{listings}
\hypersetup{
bookmarksnumbered=true,
bookmarksopen=true,
bookmarksopenlevel=1,
}
\ifCLASSINFOpdf
\else
\fi
\usepackage{amsmath}
\usepackage{url}

\begin{document}
%
\title{Uncovering Gaps Between RFC Updates and TCP/IP Implementations: LLM-Facilitated Differential Checks on Intermediate Representations}

\author{\IEEEauthorblockN{Yifan Wu\textsuperscript{†}, Xuewei Feng\textsuperscript{‡}, Yuxiang Yang\textsuperscript{‡}, Ke Xu\textsuperscript{‡*}}
\IEEEauthorblockA{\textsuperscript{†}Peking University\\
\textsuperscript{‡}Department of Computer Science and Technology \& BNRist, Tsinghua University\\
\textsuperscript{*}Zhongguancun Lab\\
mir4c1e512@gmail.com, 
fengxw06@126.com, 
yangyx22@mails.tsinghua.edu.cn, xuke@tsinghua.edu.cn
}
}



%


\maketitle

\begin{abstract}

As the core of the Internet infrastructure, the TCP/IP protocol stack undertakes the task of network data transmission. 
However, due to the complexity of the protocol and the uncertainty of cross-layer interaction, there are often inconsistencies between the implementation of the protocol stack code and the RFC standard. 
This inconsistency may not only lead to differences in protocol functions but also cause serious security vulnerabilities. 
At present, with the continuous expansion of protocol stack functions and the rapid iteration of RFC documents, it is increasingly important to detect and fix these inconsistencies.
With the rise of large language models, researchers have begun to explore how to extract protocol specifications from RFC documents through these models, including protocol stack modeling, state machine extraction, text ambiguity analysis, and other related content.
However, existing methods rely on predefined patterns or rule-based approaches that fail to generalize across different protocol specifications. 
Automated and scalable detection of these inconsistencies remains a significant challenge.
In this study, we propose an automated analysis framework based on LLM and differential models. 
By modeling the iterative relationship of the protocol and based on the iterative update relationship of the RFC standard, we perform incremental code function analysis on different versions of kernel code implementations to automatically perform code detection and vulnerability analysis.
We conduct extensive evaluations to validate the effectiveness of our framework, demonstrating its effectiveness in identifying potential vulnerabilities caused by RFC code inconsistencies. 
Our experiments reveal 15 inconsistencies between code implementations and protocol specifications, including ISN generation, TCP challenge acknowledgments, TCP authentication, and TCP timestamp options across multiple operating systems. These inconsistencies can introduce serious vulnerabilities (e.g., traffic amplification and replay attacks) in the TCP/IP protocol suite.
These results confirm that our approach provides a practical and scalable solution for a problem previously reliant on manual expertise.

\end{abstract}


%
\IEEEpeerreviewmaketitle

\section{Introduction}
In the field of network and distributed systems, adherence to RFC (Request for Comments) specifications is crucial for ensuring the security and robustness of protocol implementations. 
The TCP/IP protocol stack, as the core of Internet infrastructure, undertakes the critical task of data transmission. 
However, inconsistencies between these specifications and their corresponding code can introduce various vulnerabilities, ranging from functional deviations to severe security risks such as traffic amplification and replay attacks.
The root causes of these discrepancies are well-known yet challenging to mitigate. RFC documents are written in natural language, inherently prone to ambiguities that complicate accurate implementation. 
Furthermore, protocol implementations often involve large and complex codebases, making manual verification both labor-intensive and error-prone. 
As network protocols rapidly evolve, the continuous issuance of RFC updates exacerbates the problem, creating a moving target for human auditors. 
This landscape creates a pressing need for automated, scalable tools to systematically bridge the gap between evolving RFC specifications and their implementations—a challenge that existing methods fail to address adequately.

Existing approaches rely on predefined patterns or rules, making them ineffective at detecting inconsistencies across diverse protocol specifications.
For example, \emph{Clang Static Analyzer}~\cite{Clang_static_analyzer} primarily focus on detecting syntactic inconsistencies but struggle with identifying functional mismatches. 
While symbolic execution~\cite{King1976} tools like KLEE~\cite{klee} and fuzzing-based methods like AFL~\cite{afl} enable deeper analysis in software systems, but suffer from path explosion and high computational costs, making them 
challenging for large-scale protocol verification.
In recent years, document-based protocol security analysis has emerged as a promising direction, leveraging NLP and LLMs to extract essential semantics from complex specification documents. Tools such as SAGE~\cite{Yen2021}, RFCNLP~\cite{Pacheco2022}, and PROSPER~\cite{Sharma2023} demonstrate the integration of NLP and formal methods to construct protocol state machines and detect ambiguities. Moreover, LLM-guided fuzzing approaches~\cite{Meng2024, Xia2024} utilize structured representations to automate diverse test generation. Document-based protocol security analysis has shown practical success in analyzing real-world systems like payment protocols~\cite{chenyi1} and the 3GPP ecosystem~\cite{chenyi2}.
%
However, a critical gap remains: the end-to-end, automated uncovering of inconsistencies directly caused by RFC updates across diverse TCP/IP implementations.
%
Although EBugDec~\cite{ebugdec} studies RFC-evolutionary bugs, it heavily depends on code annotations and standardized version management.
Consequently, the field still lacks a generalized, automated solution, leaving consistency verification largely a manual and unscalable process.


%

In this paper, we propose a novel, automated framework that fundamentally shifts how consistency between RFC specifications and protocol stack code is verified. Our approach synergistically combines large language models and knowledge graphs with differential checking techniques to systematically detect and highlight inconsistencies. 
The core of our method involves two key phases: first, leveraging LLMs to extract protocol-relevant entities from both RFCs and code, modeling a structured correlation between specifications and implementations; second, modeling protocol evolution by constructing an update relationship graph to enable precise, incremental differential analysis.

To evaluate the effectiveness of our proposed framework, we conducted an extensive empirical study. 
We applied our framework to 8 RFCs and their implementations across 7 versions of major operating systems, including Linux, FreeBSD, OpenBSD, NetBSD and Android. 
Additionally, we utilize state-of-the-art LLMs, such as GPT-4o~\cite{gpt4o} and DeepSeek-V3~\cite{Liu2024}, to facilitate automated extraction and verification.
Experimental results show that our approach achieves 91.1\% accuracy and an F1 score of 0.857 based on GPT-4o, significantly outperforms vanilla LLM-based detection.
As a result, our framework identified 15 inconsistencies between the code implementations and protocol specifications, including ISN generation, TCP challenge acknowledgment, TCP authentication, and TCP timestamp options, , which can introduce serious vulnerabilities like traffic amplification, data injection, and TCP RST spoofing.
%

\noindent \textbf{Contributions}. Our main contributions are as follows:
\begin{itemize}
[leftmargin=*]
\item We propose a novel framework that integrates knowledge graphs and differential checking to systematically verify consistency between RFC specifications and real-world protocol implementations.

  \item We implement our framework based on a hybrid analysis pipeline.
  The implementation is in Python, with default support for DeepSeek and compatibility with GPT series models.
  We will open source our prototype implementation to foster further research in network security.

\item We conduct extensive evaluations of our framework, demonstrating its effectiveness by automatically identifying 15 inconsistencies across 8 RFCs and 7 kernel versions, thereby providing a validated solution for a problem previously reliant on manual effort.
\end{itemize}

\section{Background}
The TCP/IP protocol stack has experienced decades of development.
As security issues and new features emerge, RFC standard documents are frequently updated, making compatibility and maintenance between versions a huge challenge.
There are significant differences in code implemented by different vendors and communities, leading to increased collaboration and interoperability issues.
At the same time, due to developers’ different understandings of standards and the fact that certain features are not implemented according to the standards (or are not implemented), inconsistencies between code and protocol standards may lead to corresponding security issues and functional failures.

\subsection{Protocol Security}
For example, a well-known example of a TCP RST attack exploits the lenient handling of RST packets defined in RFC 793~\cite{rfc793}.
According to RFC 793, as long as the sequence number of a forged RST packet falls within the TCP sliding window of the receiver, the connection is immediately closed.
An attacker can easily disconnect a legitimate TCP connection by guessing the connection's four-tuple (source IP, destination IP, source port, destination port) and the sliding window range, and then sending a forged RST packet.
Such attacks can lead to service disruptions (e.g., termination of HTTP, SSH, or database connections) and present a denial of service (DoS) risk, particularly affecting long-lived connections like video streams or remote control services.
RFC 5961~\cite{rfc5961} addresses this issue by introducing stricter sequence number validation and a challenge ACK mechanism, which enhances the robustness of TCP and effectively mitigates blind RST attacks.
However, if these mechanisms are not implemented, the system remains vulnerable to such attacks.


\subsection{Related Work}

\subsubsection{Code Generation and Consistency Verification}

In recent years, the rapid development of artificial intelligence~\cite{Turing1950}, particularly deep learning~\cite{LeCun2015} and Transformer~\cite{Vaswani2017}-based architectures, has led to significant breakthroughs in automated code generation. 
Pre-trained large language models, initially developed for NLP, have been adapted for coding tasks. 
Early models based on BERT~\cite{Devlin2018} (e.g., CuBert~\cite{Kanade2019}) have paved the way for more advanced systems like CodeX~\cite{Chen2021}, AlphaCode~\cite{Li2022}, and CodeGeeX~\cite{Zheng2023}.

Ensuring that generated code conforms to specifications is equally critical. 
Current research focuses on three main verification approaches. 
\begin{itemize}
\item Test case-based method utilizes standardized languages such as TTCN-3~\cite{Grabowski2003} to design and automate rigorous testing. 
\item Symbolic execution~\cite{King1976} replaces concrete values with symbolic ones to systematically explore all execution paths, with enhancements through fuzzing~\cite{Stephens2016} and machine learning~\cite{Li2016} to address challenges like path explosion.
\item Differential analysis~\cite{McKeeman1998} compares behaviors across different implementations or versions, including analyses based on control flow graphs and finite state machines~\cite{Pacheco2022} to identify discrepancies. 
\end{itemize}  

\subsubsection{Multi-step Reasoning in Large Language Models}
Large language models (e.g., GPT-4~\cite{Achiam2023}, DeepSeek) leverage Transformer-based pre-training and Self-Attention mechanisms~\cite{Vaswani2017} to learn rich language patterns and reasoning abilities. 
To handle complex tasks like protocol analysis and code understanding, these models employ multi-step reasoning strategies. 
For instance, Chain-of-Thought (CoT)\cite{Wei2022} guides the model through intermediate reasoning steps—whether via few-shot examples or zero-shot prompts (e.g., "let's think step by step"\cite{Kojima2022}). 
Forest-of-Thought (FoT)\cite{Bi2024} builds on the Tree-of-Thought framework\cite{Yao2024} by integrating multiple reasoning trees with dynamic path selection and self-correction, enhancing both efficiency and accuracy. 
Least-to-Most (L2M)~\cite{Zhou2022} decomposes complex problems into sequential sub-problems, enabling improved generalization and problem-solving. 
These techniques collectively underpin the advanced multi-step reasoning capabilities that are central to modern LLMs.

\subsubsection{Knowledge Graph-Based Retrieval-Augmented Generation}
Retrieval-Augmented Generation (RAG) enhances language model outputs by integrating external knowledge, retrieving pertinent documents via a vector database to provide real-time contextual input~\cite{Lewis2020, Khandelwal2019}. 
This approach not only mitigates hallucination but also improves accuracy, though it faces challenges in multi-step inference and capturing complex semantic relationships~\cite{Zhao2024}. 
Incorporating Knowledge Graphs (KG)\cite{KGwiki}—which structure information as interconnected entities and relationships—further enriches RAG by transforming unstructured text into structured semantic data. 
This integration has boosted performance in applications such as question-answering, semantic search, and recommendation systems\cite{Liu2020, Yao2019}, marking a promising direction for enhancing the diversity and precision of generated content.

\subsubsection{Contrast with Formal Methods}
Our framework builds upon but fundamentally diverges from traditional formal verification techniques such as rule-based static analysis and model checking. The key distinction lies in the trade-off between automation and formal guarantee.
Formal methods excel in providing mathematical rigor and reliability for well-defined properties once a precise model is constructed—a process that is often manual, labor-intensive, and requires deep expertise. However, this strength is also their primary limitation, making them unable to scale with the rapid evolution of protocols and vulnerable to missing inconsistencies in clauses that were not explicitly modeled.
In contrast, our LLM-driven framework prioritizes end-to-end automation and generalization. Unlike traditional static analyzers that rely on manually crafted rule sets, it directly consumes RFC documents and kernel code, enabling semantic reasoning over natural-language specifications — a capability fundamentally absent in conventional static analysis.

We acknowledge the trade-offs. Our approach lacks the logical guarantees of formal verification and inherits uncertainties from LLM behavior. Nonetheless, we view these paradigms as complementary: our framework provides scalable, automated exploration, while formal methods can offer rigorous validation of the critical components it identifies—a promising direction for hybrid analysis pipelines.

\section{Motivation}


The process of ensuring consistency between RFC specifications and protocol stack code implementations is inherently complex and challenging. 
This complexity arises from the ambiguity and non-standard nature of RFC documents, the intricate and lengthy structure of protocol stack code, and the semantic gap between specification descriptions and real-world implementations. 
Addressing these challenges requires overcoming several key obstacles, the main challenges will be discussed in detail below.

\textbf{Diversity Across Implementations}
Multiple independent implementations of the same protocol are maintained by different teams or organizations, often with significant differences in coding style and design choices. 
Detecting inconsistencies between different implementations while following a unified RFC increases the complexity of the detection process.

\textbf{Scalability in Large-Scale Code and Specifications}
The amount of code implemented in the protocol stack and the length of the RFC documents are very large, which poses a serious challenge to the scalability of the detection process. 
In addition, there are multiple protocol implementation versions (such as Linux and FreeBSD), and achieving comprehensive coverage requires a lot of human effort and computing resources, resulting in high time costs and computing costs.

\textbf{Mismatch Between Specification Abstraction and Implementation Details}
RFC documents often focus on protocol specifications at a high level, emphasizing functional behavior rather than the concrete implementation details found in code. 
This abstraction creates a semantic gap between RFCs and code, complicating the process of identifying inconsistencies and ensuring functional equivalence.

While these challenges highlight the complexities of aligning protocol specifications with real-world implementations, an effective solution requires addressing fundamental research questions. 
Specifically, our approach aims to address several critical research questions: 
\begin{itemize} 
\item \textbf{RQ1: How can we effectively correlate specifications with code?} 
\item \textbf{RQ2: How can we efficiently perform large-scale code consistency checking tasks?} 
\item \textbf{RQ3: How can we bridge the semantic gap between specifications and code?} 
\end{itemize}

\begin{figure}[htbp]
  \centering
  \includegraphics[width=0.47\textwidth]{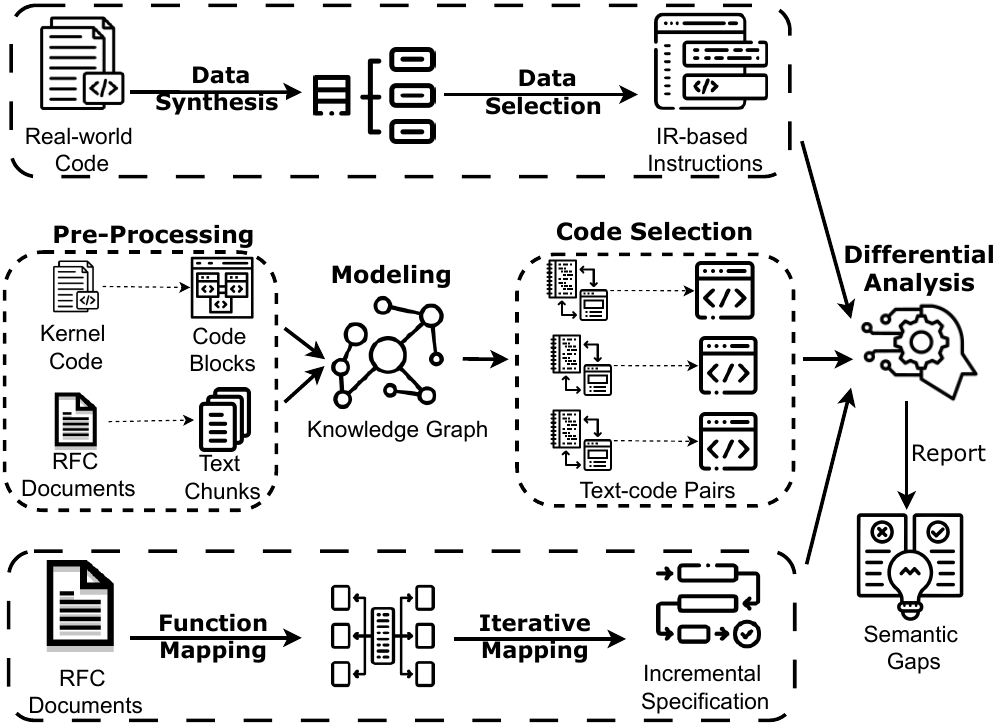}
  \caption{Differential Analysis Framework} 
  \label{fig:differential_analysis_framework} 
  \end{figure}

\section{Methodology}

This section introduces the design of our analysis framework, aiming at the consistency detection of protocol specification and protocol stack code implementation from the perspective of incremental updates.

\subsection{Overview}
The analytical framework is illustrated in Figure~\ref{fig:differential_analysis_framework}, designed to systematically verify the consistency between TCP/IP protocol specifications and their corresponding code implementations, thereby uncovering potential security vulnerabilities and deviations in implementation. 
Leveraging an automated workflow that encompasses data preprocessing, information extraction, and modeling, data synthesis, and differential analysis, this approach efficiently extracts critical protocol-related information from massive RFC documents and extensive kernel codebases. 
By constructing knowledge graphs and iterative update relationship graphs, we precisely capture incremental changes across different versions of protocol specifications, enabling accurate consistency verification between the protocol standard and its implementation.

Specifically, the data preprocessing module standardizes the format of raw protocol documents and code by cleansing redundant information and partitioning the content into manageable blocks with clearly marked contextual boundaries. 
The information extraction and modeling module then employs pre-trained language models and knowledge graph techniques to extract core entities—such as protocol states, events, and actions—from both text and code, thereby establishing a mapping between the protocol specifications and code implementations. 
Finally, the differential analysis module utilizes the differential triplet dataset we synthesised to conduct an in-depth comparison of the incremental changes in protocol specifications across versions, ultimately detecting deviations in the code implementation. 

This method offers three primary advantages:
\begin{itemize}
\item \textbf{Low Computational Overhead}: By leveraging knowledge graph-based correlation modeling and block segmentation, we effectively narrows the analysis scope and reduces redundant data processing, significantly lowering the consumption of computational resources.
\item \textbf{High Scalability}: The incremental update mechanism and iterative mapping based on protocol functionality enable our method to adapt to the continuous evolution of protocol specifications and dynamic changes in large-scale codebases, making it suitable for multi-version and multi-protocol scenarios.
\item \textbf{High Detection Accuracy}: The intermediate representation based differential analysis strategy effectively bridges the semantic gap between protocol specifications and code implementations, allowing for the precise detection of incremental changes and inconsistencies, thereby significantly enhancing the accuracy of vulnerability detection.
\end{itemize}

\subsection{Data Preprocessing}
This phase encompasses RFC documents and kernel codes, both undergoing tailored preprocessing steps to optimize analytical efficiency. 
The following outlines the key strategies applied to each data source:

\textbf{Document Parsing}: 
RFC documents are an important source of Internet protocol standards,
often contain multidimensional content such as tables of contents, footnotes, authors, appendices, and other metadata. 
Although these contents are of reference value to human readers, they are irrelevant to the specification definition
 and may introduce noises, thus affecting the accuracy of the model. 
Therefore, this method first cleans the RFC document to remove the above irrelevant parts and ensures that only the core content related to the protocol is retained. 
Then, the cleaned document is segmented twice according to the text structure such as chapter and paragraph to maintain the hierarchical structure of the document.
Notably, the ASCII art diagram in the document is extracted separately to facilitate the subsequent semantic analysis and information extraction.

\textbf{Kernel Code Parsing}: 
Kernel code is frequently extensive and intricate, encompassing numerous macro definitions, inline functions, and inter-file references. 
Direct analysis can lead to difficulties in comprehension, poor readability, and other challenges. 
Given the multitude of cross-file references within kernel code, directly parsing all header files and predefined content may complicate the analysis process and increase the complexity of understanding (for instance, the "tcp.c" file in Linux include over 40 files directly). 
Consequently, we establish an appropriate level of abstraction, concentrating on the upper-layer implementation of network protocols while avoiding an in-depth examination of lower-level common kernel details, 
 simplifying the analysis process and highlights key logic. 
First, our framework extracts the network protocol module code from the kernel code. 
Second, our framework analyzes the syntax tree of this extracted code at an appropriate level of abstraction to identify critical information such as function definitions, parameters, bodies, and comments, 
  transforming code into a structured format.


\begin{figure}[htbp]
  \centering
  \includegraphics[width=0.47\textwidth]{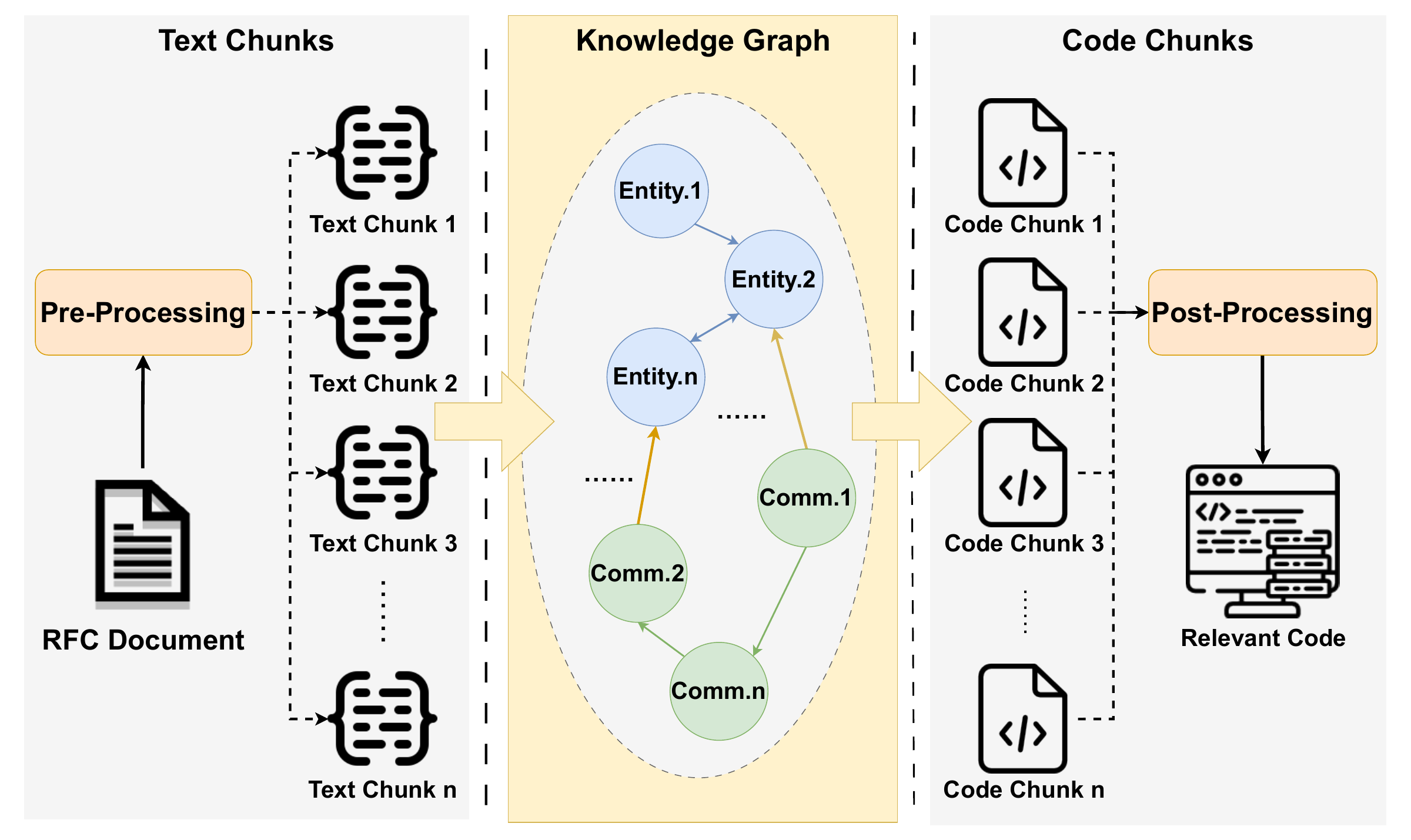}
  \caption{Knowledge Graph-Driven Information Extraction and Modeling} 
  \label{fig:knowledge_graph} 
  \end{figure}

\subsection{Information extraction and modeling}
\label{sec:information_extraction}

In this phase, we utilize a knowledge graph to construct the relationship between these elements, thereby enhancing the efficiency of achieving the consistency verification goals set forth by this method. 
This section will give detailed explanations in three subsections: cross-domain semantic alignment, incremental specification graph construction, and code semantic restoration.

\subsubsection{Cross-Domain Semantic Alignment}

Our methodology employs a knowledge graph-centric framework to bridge the semantic gap between protocol specifications and code implementations through structured semantic alignment. As shown in Figure ~\ref{fig:knowledge_graph}, this process consists of three phases: text and code chunking, knowledge graph construction, and code chunks retrieval, enabling fine-grained mapping from RFC text chunks to kernel code functions.

In real-world analysis scenarios, RFC documents and kernel code often exhibit extensive length and intricate structures.         
Directly analyzing such large-scale documents and codebases frequently leads to computational resource bottlenecks and efficiency challenges.         
Therefore, the alignment begins with decomposing protocol specifications and kernel code into smaller chunks.         
Then we use pre-trained language models to extract protocol-centric entities from both text and code chunks, including protocol states, transition events, and actions, organizing them into a hierarchical knowledge graph within a shared semantic space.

To retrieve relevant code chunks, our framework use a dual-path retrieval strategy, 
combining direct entity mappings with community-based clustering to identify all associated kernel code fragments.  



\subsubsection{Incremental Specification Graph Construction}
\begin{figure}[!t]
  \centering
  \includegraphics[width=3.4in]{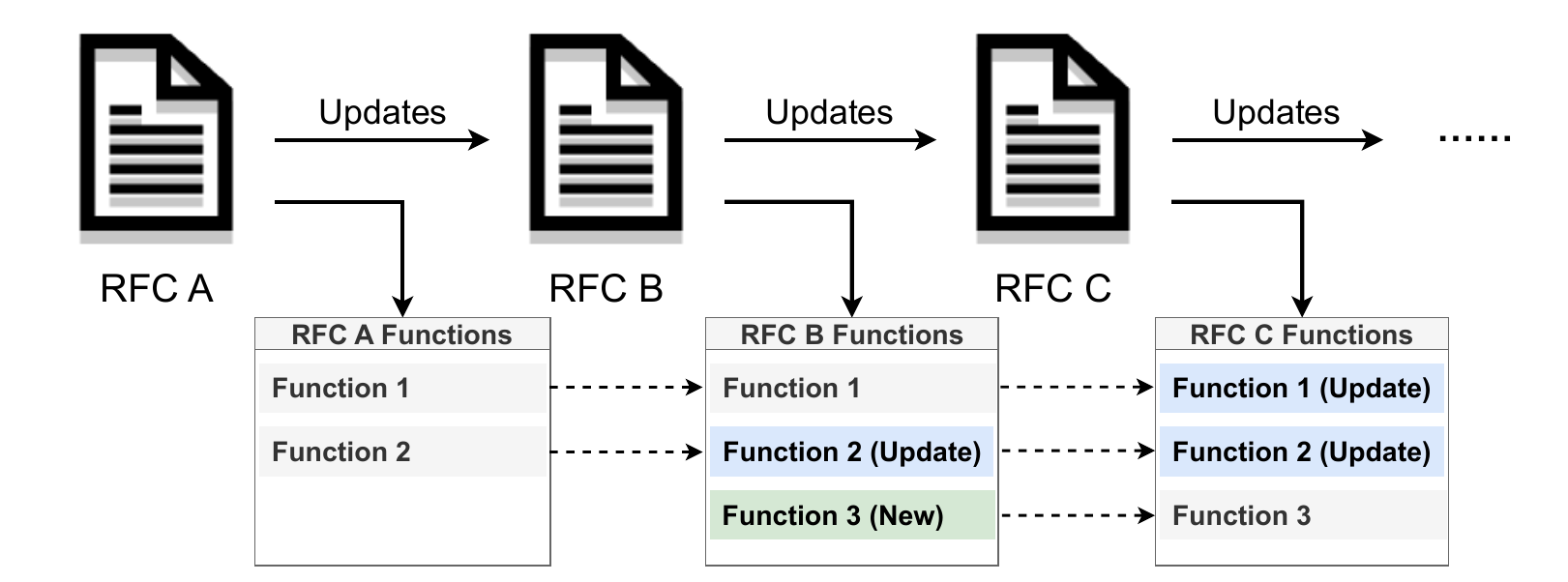}
  \caption{RFC Functional Mapping and Iterative Relationships} 
  \label{fig:rfc_func_updates} 
  \end{figure}

The evolution of protocol specifications is inherently iterative. 
With advancements in technology and evolving security requirements, protocol specifications are continually updated—typically through revisions and deprecations in RFC documents, which are indicated in the Standard Track to guide developers in their code implementations. 
However, system development and standard updates are not always synchronized, therefore different versions of systems may not timely adapt to the latest standards, leading to inconsistencies between the code implementation and the specification, potentially introducing security vulnerabilities.

To accurately capture the evolution of protocol functionalities and ensure that code implementations reflect the latest specification changes, the proposed method focuses on modeling the incremental updates within the protocol specifications by constructing an iterative update graph based on protocol functionalities. 
This process is divided into two key steps:
\begin{itemize}
\item 
First, each RFC document is subjected to functional extraction. 
RFC documents typically exhibit a highly structured hierarchical organization, divided into core sections such as Protocol Overview, Protocol Functionality, Security Analysis, and Conclusion. 
A complete protocol functionality is usually concentrated within a specific section and includes critical components such as state machine definitions, interaction rules, and boundary conditions. 
In the data preprocessing stage, the RFC documents are segmented according to their section structure, and then a pre-trained language model is employed to deeply parse each segmented section, automatically extracting key concepts such as protocol states, events, and interaction rules, thereby generating standardized functional entries.
\item
Next, based on these extracted functional entries, we construct an update-based relationship graph among RFC documents. 
This graph is built by analyzing version iteration information derived from the RFC document metadata and revision notes, organizing related RFC documents in sequential order. 
As a result, it clearly identifies which functionalities have been updated or modified in the new versions and which functionalities were already defined in earlier versions. 
The comparison of functionalities across versions establishes a mapping that forms an iterative update chain of protocol functions.
\end{itemize}

Finally, the above steps are integrated to construct a unified incremental specification graph as shown in Figure ~\ref{fig:rfc_func_updates}, which represents protocol evolution at the granularity of functional entries. 
This graph visually presents the inheritance, updates, and deprecation of protocol functionalities across different RFC versions.
By employing this incremental update modeling strategy based on protocol functionalities, we can efficiently handle large volumes of RFC documents,  reduce the computational complexity and time overhead of full-scale comparisons while enhancing the precision and efficiency of consistency detection.

\subsubsection{Code Semantic Restoration}
\begin{figure}[!t]
  \centering
  \includegraphics[width=2.3in]{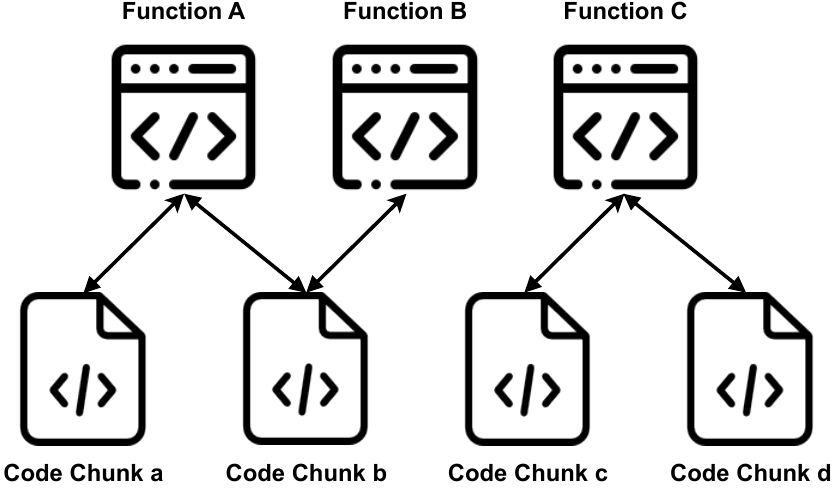}
  \caption{Kernel Code Chunk-to-Function Mapping} 
  \label{fig:chunk_to_func} 
  \end{figure}
Traditional fixed-length segmentation methods often lead to semantic fragmentation. 
Although segmenting code by function is an alternative, differences in function lengths result in long functions naturally receiving higher weights, while short functions may be undervalued, complicating global weight normalization. 
To address these issues, we employ a hybrid approach that combines block segmentation with function-level mapping, thus preserving overall processing efficiency while restoring function-level semantics via AST(abstract syntax tree) parsing.

Initially, the preprocessed code is divided into fixed-length chunks of 500 tokens, with an overlapping redundancy of approximately 10\% introduced between chunks to preserve semantic continuity at the boundaries. 
Specifically, the chunk boundary is determined by aligning the chunk's end with the end of the current function whenever possible; if a complete function end cannot be achieved in redundancy, the chunk is terminated at the nearest end-of-statement. 
This approach minimizes the risk of severing critical logic.

Subsequently, we use AST parsing to extract key elements, including function definitions, parameters, function bodies, and annotations, establishing a mapping between functions and their corresponding chunks. 
By recording the start and end positions of each function, we can accurately reconstruct functions that span multiple chunks. 
Given that a single chunk may contain multiple functions and a function might span several chunks, this process ensures the establishment of a bidirectional and traceable mapping between code chunks and functions as illustrated in Figure ~\ref{fig:chunk_to_func}.
By reconstructing the logical structure of code,  our method overcomes the semantic fragmentation inherent in fixed segmentation and ensures consistency in global weight calculation.

\subsection{Differential Analysis Model}
\label{sec:differential_analysis_model}
Ensuring consistency between the protocol specification and the code implementation is a critical but challenging task. 
This is mainly due to the complexity and scale of the protocol specification and code implementation. Traditional full-coverage analysis methods are often faced with high computational cost, low efficiency, and limited accuracy, which makes it difficult to be applied in dealing with large-scale RFC updates and code bases.
To solve these problems, we introduce a differential analysis framework based on incremental validation.

\subsubsection{Incremental Verification Based on Specification Updates}

\begin{figure}[!t]
  \centering
  \includegraphics[width=3.49in]{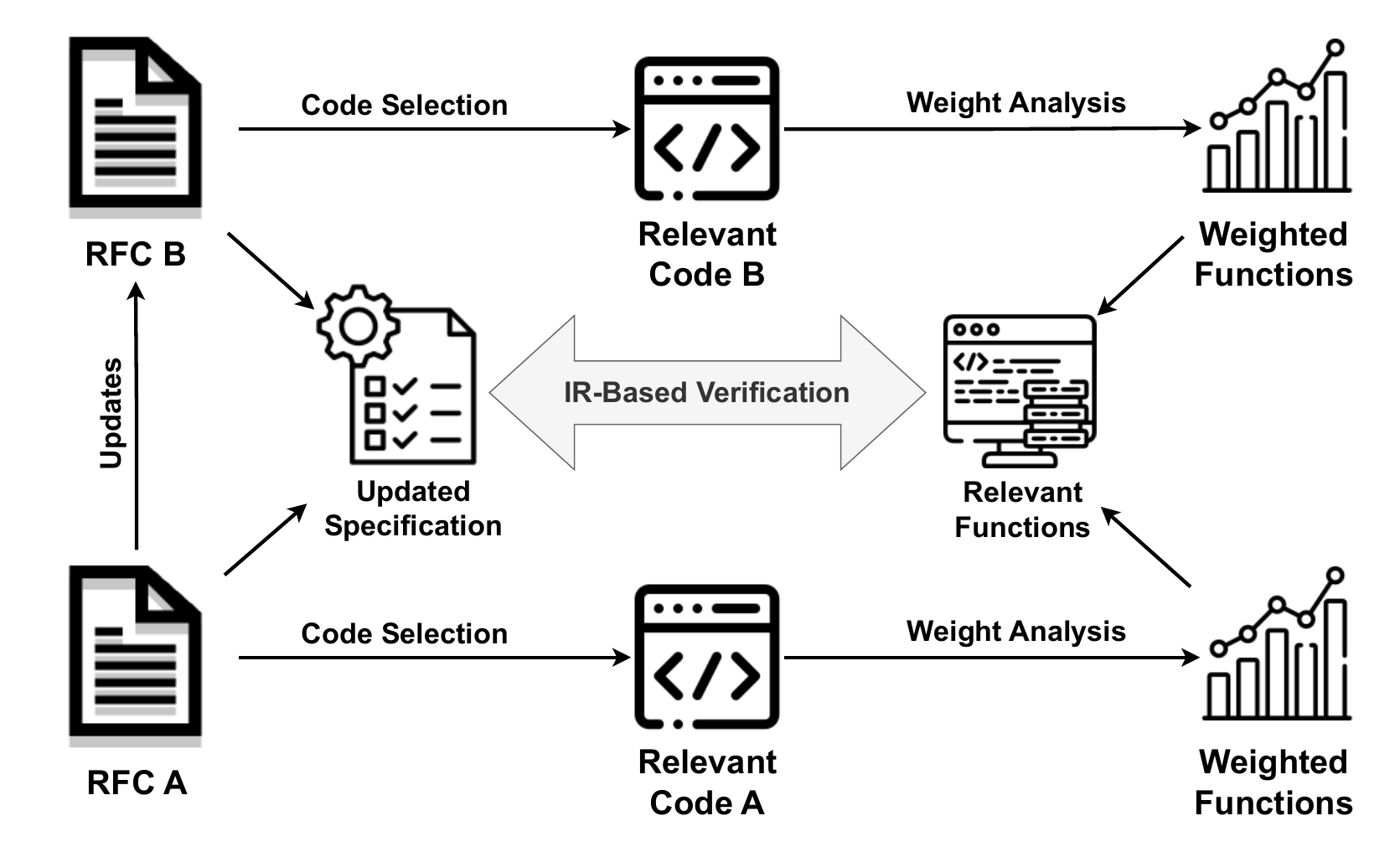}
  \caption{Incremental Verification Based on Protocol Specification Updates} 
  \label{fig:update_relevant_code} 
  \end{figure}

  Traditional consistency verification methods typically rely on full-scale comparisons to detect discrepancies between code implementations and protocol specifications. 
However, with the rapid development of Internet technologies, the scale of RFC documents and protocol stack codebases has increased dramatically, far exceeding the processing capabilities of conventional methods. 
Conducting a full-scale analysis on such massive datasets incurs high computational costs and significantly reduces efficiency, making it unsuitable for real-world engineering scenarios. 
To address this challenge, we introduce an incremental verification mechanism as shown in Figure ~\ref{fig:update_relevant_code}, which focuses on the update relationship between RFC documents and their functionality, thereby pinpointing the key change areas for targeted consistency checks.

Specifically, we first perform functional extraction on each RFC document by employing pre-trained language models to extract core elements of protocol functionality—such as protocol states, events, and interaction rules, and standardize them into functional entries.
Next, we construct an evolution graph of RFC documents with their functionality to clearly delineate the changes between versions, including additions, modifications, and deletions of functionalities.     
Based on this iterative update graph, the method conducts consistency verification exclusively on the updated segments rather than performing a full-scale comparison of the entire codebase.
By mapping the updated functional entries to their corresponding segments in the code implementation, the system effectively narrows the detection scope, reduces redundant comparisons, and thereby significantly improves both the efficiency and accuracy of the consistency analysis.


\subsubsection{Consistency Verification Based on Intermediate Representations}
\begin{figure}[!t]
  \centering
  \includegraphics[width=3.49in]{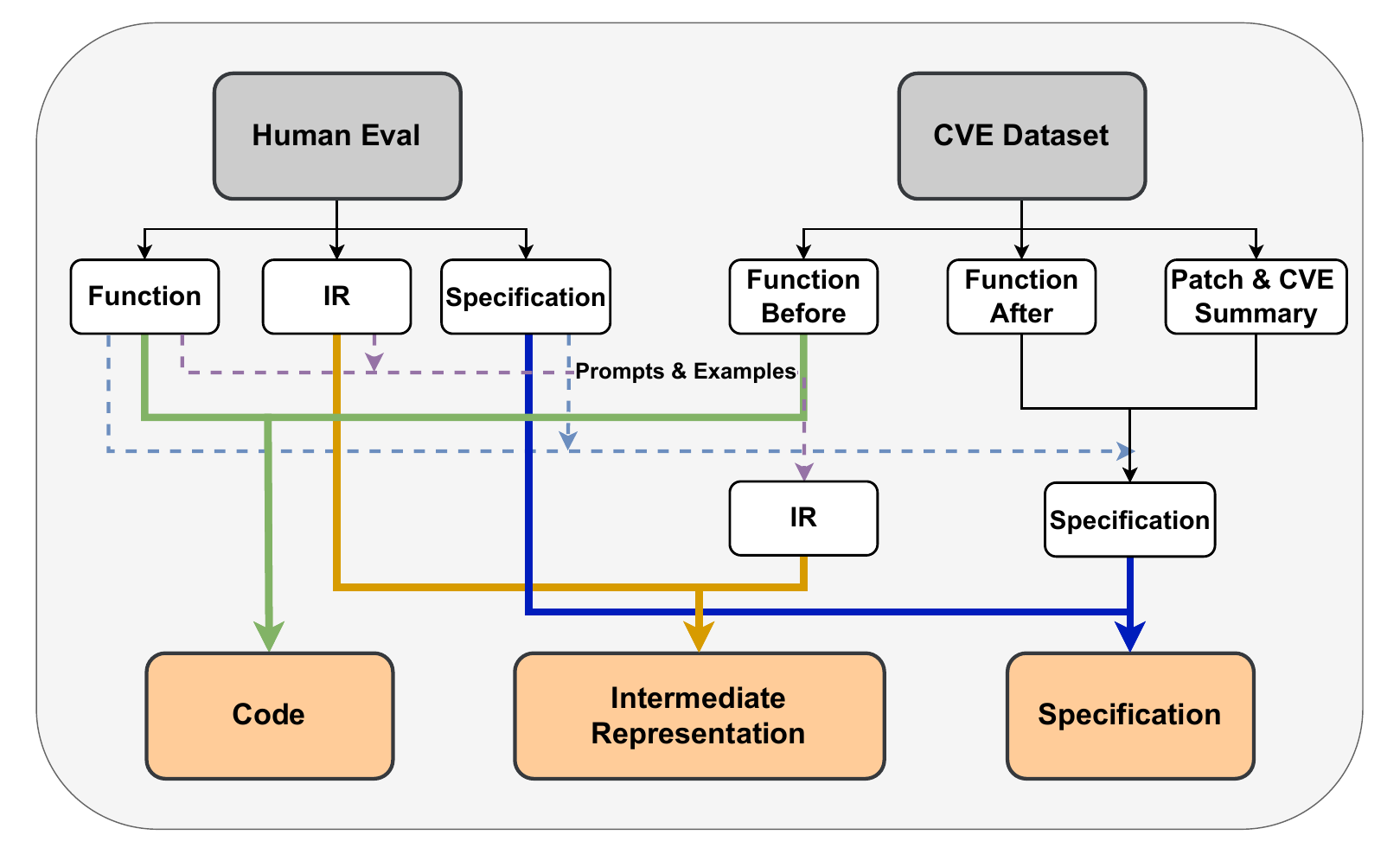}
  \caption{Data Synthesis Workflow} 
  \label{fig:data_synthesis} 
  \end{figure}
A major challenge in verifying consistency between protocol specifications and code implementations lies in the semantic gap: protocol specifications are typically expressed in highly abstract natural language, detailing design principles, message formats, and state transition rules, while code implementations capture concrete, low-level logic and details.
This disparity in abstraction levels and semantic expression makes direct comparison difficult, often leading to misdetections or omissions.  To address this challenge, it is essential to develop a method that bridges the semantic gap between specifications and implementations, enabling high-precision consistency verification.

Therefore, we introduce a consistency verification strategy based on intermediate representations and construct a differential triplet dataset that maps together the protocol specification text, an intermediate representation, and the corresponding source code.
The "intermediate representation" here is defined as a natural language-based code description that directly guides the implementation of functional code, thereby serving as a semantic alignment bridge between the abstract specification and the concrete implementation.
Specifically, the positive samples are derived from Human Eval~\cite{Chen2021} by extracting function descriptions to generate intermediate representations, while the negative samples are constructed from CVE Dataset~\cite{Fan2020} by capturing discrepancies between before-patch and after-patch code along with the CVE summary. 
The workflow of data synthesis is shown as Figure ~\ref{fig:data_synthesis}.
We synthesize a total of 330 entries of triplet data for the experiment, which enables the model to learn the key correspondences between specifications and implementations.

During differential analysis, the updated specification and selected code are fed into the model separately, with the intermediate representation serving as a semantic bridge for alignment and comparison. 
Specifically, we calculates semantic similarity and selects the top K most similar examples from both positive and negative samples, sorting them in ascending order of complexity, to serve as guiding examples for the model to generate the corresponding intermediate representations. 
This process automatically detects inconsistencies between the specification and the implementation, putting forward a feasible solution to this challenge.

\subsection{Method Advantages}
By employing an incremental verification mechanism along with differential analysis based on intermediate representations, our method effectively reduces the computational complexity while ensuring high detection precision, showing advantages in terms of computational efficiency, scalability, and detection accuracy.

\subsubsection{Low Computational Overhead}
Assume the total scale of RFC documents is $N$ with an average length of $Len_{RFC}$, and the codebase size is $M$. 
Traditional full-scale comparison methods typically analyze the entire code for each RFC, resulting in a worst-case computational complexity of:
\begin{align}
  O(N \cdot (Len_{RFC}+M))
\end{align}

In real-world scenarios, where both $N$ and $M$ are large, this approach incurs enormous computational overhead. 
In contrast, our method focuses solely on the updated portions of RFC documents and their corresponding regions in the code. 
Let $\Delta Len_{RFC}$ denote the incremental size of RFC functional updates and $\Delta M$ the affected code size. 
The incremental verification approach achieves a complexity of:
\begin{align}
  O(N \cdot (\Delta Len_{RFC}+\Delta M))
\end{align}

Since typically $\Delta Len_{RFC} \ll Len_{RFC}$ and $\Delta M \ll M$, the overall computation is greatly reduced. 
Moreover, analyzing costs from a token perspective (commonly used as a billing unit in large model API services) further demonstrates the efficiency gains. 
In full-scale analysis, the total token consumption is:
\begin{align}
  Token_{\text{Na\"{i}ve}} = N \cdot (Len_{RFC}+M)
\end{align}
While in our method, the token consumption in reasoning is:
\begin{align}
  Token_{\text{Reasoning}} = N \cdot (\Delta Len_{RFC}+\Delta M)
\end{align}
An additional one-time token cost is incurred during knowledge graph construction:
\begin{align}
  Token_{\text{Graph}} = N \cdot Len_{RFC}+M
\end{align}
Thus, the total token consumption of our method is:
\begin{align}
  Token_{\text{Total}} &= Token_{\text{Reasoning}} + Token_{\text{Graph}}\nonumber \\
  &= N \cdot (\Delta Len_{RFC}+\Delta M) + N \cdot Len_{RFC}+M
\end{align}
The difference in token consumption compared to full-scale analysis is:
\begin{align}
  \Delta Token &= Token_{\text{Na\"{i}ve}} - Token_{\text{Total}}\nonumber \\
  &= (N-1) \cdot M - N \cdot (\Delta Len_{RFC}+\Delta M)
\end{align}
When the total number of RFC documents is large, it can be approximated as:
\begin{align}
  \Delta Token \approx  N \cdot (M - \Delta Len_{RFC}+\Delta M)
\end{align}
In real-world scenarios, the size of incremental updates ($\Delta Len_{RFC}$ and $\Delta M$) is only a small fraction of the codebases. 
Consequently, our method reduces the computational complexity, resulting in significant savings in both time and resources.

\subsubsection{High Scalability}

Our method effectively handles massive RFC documents and large codebases by combining block segmentation with knowledge graph-based correlation modeling.    
For instance, as modern operating systems evolve, the code volume increases dramatically—for example, the total code lines in the /net/ipv4 directory in Linux grew from 84k in version 3.6 to 113K in version 6.9.    
While current LLM-based code generation and detection techniques are powerful, they are limited by context window sizes (typically 32K, 64K, or 128K tokens) and cannot process entire codebases in a single session.    
By using segmentation and focused analysis on the updated regions extracted from the RFC documents, our method avoids redundant processing, thereby substantially improving scalability and making it applicable to multi-version and multi-protocol scenarios.

\subsubsection{High Detection Accuracy}

Traditional full-scale analysis often struggles with the semantic gap between the abstract protocol specifications and the concrete code implementations, leading to misdetections.      
Therefore, our method leverages an incremental update verification strategy in conjunction with differential analysis based on intermediate representations.     
By aligning incremental updates with the corresponding code via a differential triplet dataset which in turn guides the LLM to generate intermediate representations that capture the differences between the updated specification and its implementation, this approach effectively bridges the semantic gap.   
Experimental results indicate that this strategy significantly improves both precision and recall by ensuring that only the truly affected regions are analyzed, thereby reducing false positives and optimizing overall detection efficiency.

\section{Implementations and Evaluations}\label{sec:results}

\subsection{Implementations}

Our framework is implemented in Python and consists of three core modules: data preprocessing, information extraction, modeling, and differential analysis. 

In the data preprocessing module, we use regular expressions to clean RFC documents by removing irrelevant content, segmenting chapters, and extracting ASCII art diagrams. 
We rely on the open-source library pycparser and a fake include mechanism for kernel code to generate an AST, avoiding deep dependency on generic kernel functions.

In the information extraction and modeling module, we draw on parts of GraphRAG, constructing a unified knowledge graph to align preprocessed RFC with kernel code and reconstruct functions from code chunks based on an AST-derived approach. 
To extract protocol functionalities, we leverage LLM(defaulting to DeepSeek) to analyze incremental updates between RFC versions, forming an Incremental Specification Graph.

In the differential analysis module, we base on the Incremental Specification Graph to verify whether new functionalities are implemented. 
Each verification retrieves the top-k most similar instances via BERT embeddings and BM25 scoring from the synthesised differential triplet dataset. 
These instances guide the LLM in generating an intermediate representation, which is then used to verify the consistency between the incremental functionality and kernel code.


\subsection{Experimental Setup}
TCP plays a central role in the protocol stack, and its functionality and security directly impact the stability and reliability of network communications.
Therefore, this study selects the TCP protocol as the primary focus.
Additionally, due to the lack of publicly available datasets suitable for large-scale consistency verification of protocol implementations and the high cost and subjectivity associated with manual analysis, the research concentrates on an in-depth examination of TCP.
This enables a rigorous evaluation of the proposed consistency detection method, providing robust theoretical and practical support for ensuring network system stability and security under limited research resources and time constraints.

In the experimental setup, we filter 144 TCP-related RFC documents and seven different versions of system kernel code as test subjects, including:
\begin{itemize}
  \item \textbf{Linux}: Versions 6.9, 3.6, and 2.6.39, where version 6.9 represents a recent release while versions 3.6 and 2.6.39 reflect long-maintained releases.
  \item \textbf{BSD}: FreeBSD 13.3, NetBSD 9.4, and OpenBSD 7.5, which are among the more current releases in their respective systems.
  \item \textbf{Android}: The 4.19-stable version, a long-term maintained stable release based on Linux 4.19.
\end{itemize}

These different versions of the kernel are not completely consistent in their functional implementation and are widely used in the real world, so examining these codes can help to fully evaluate the effectiveness of our methods.

Additionally, due to the lack of publicly available datasets suitable for large-scale consistency verification of protocol implementations and the extremely high cost of analysis of every specification implementation across all code versions by human experts, we select eight representative RFC documents from four update chains as benchmarks, including:
\begin{itemize}
  \item \textbf {RFC793 → RFC1948 → RFC6528}:
  Focus on improvements in TCP initial sequence number (ISN) generation to prevent sequence number prediction attacks.
  \item \textbf {RFC793 → RFC5961}:
  Deals with TCP challenge acknowledgment mechanisms aimed at defending against TCP RST attacks, forged SYN attacks, and outdated sequence number guessing.
  \item \textbf {RFC793 → RFC2385 → RFC5925}:
  Addresses TCP authentication, transitioning from TCP MD5 signatures to TCP Authentication Options (TCP-AO).
  \item \textbf {RFC793 → RFC1323 → RFC7323}:
  Pertains to TCP performance extensions, such as window scaling and timestamps, and includes security considerations like PAWS and timestamp-related issues.
\end{itemize}

After conducting detailed analyses of the corresponding protocol implementations in each code version, we identified 56 data points to serve as the ground truth for our evaluation.

\subsection{Evaluation Metrics}

To comprehensively evaluate the proposed method's effectiveness, efficiency, and cost-effectiveness in detecting inconsistencies between protocol specifications and code implementations, we use a set of metrics.   
The evaluation metrics including \textbf {Accuracy}, \textbf {Recall}, \textbf {Recall}, \textbf {F1 Score}, \textbf {Total Token Consumption}, \textbf {Total Cost} and \textbf {Total Time}. Noting that runtime is provided as a reference due to potential variability from external factors.

\subsection{Experimental Results}
\begin{table*}[!htbp]
  \centering  
  \caption{Code Extraction Experimental Results Based on Knowledge Graph Modeling}\label{tab:coderelextract}
  \fontsize{10pt}{12pt}\selectfont
  \resizebox{0.7\textwidth}{!}{
    \begin{tabularx}{0.9\textwidth}{c*{7}{>{\centering\arraybackslash}X}}
    \toprule
      \multirow{2}{*}{\textbf{Metric}} & \multicolumn{7}{c}{\textbf{Code Version}} \\
      \cmidrule(lr){2-8}
      & \makecell{\textbf{Linux}\\\textbf{6.9}}  & \makecell{\textbf{Linux}\\\textbf{3.6}} & \makecell{\textbf{Linux}\\\textbf{2.6.39}} & \makecell{\textbf{Android}\\\textbf{4.19}} & \makecell{\textbf{FreeBSD}\\\textbf{13.3}} & \makecell{\textbf{NetBSD}\\\textbf{9.4}} & \makecell{\textbf{OpenBSD}\\\textbf{7.5}}\\
      \midrule
      TF & 1250 & 756 & 715 & 1076 & 619 & 363 & 134 \\
      TL & 37327 & 18917 & 18231 & 25787 & 27793 & 15883 & 8484 \\
      SF & 69.8 & 36.6 & 9.5 & 38.2 & 30.8 & 24.2 & 11.7 \\
      SL & 5628.7 & 1699.5 & 655.2 & 1590.3 & 3628.8 & 2769.1 & 2317.7 \\
      FER (\%) & 5.6 & 5.1 & 1.3 & 3.6 & 5.0 & 6.7 & 8.7 \\
      LER (\%) & 15.1 & 9.3 & 3.5 & 6.2 & 13.1 & 17.4 & 27.3 \\
      \bottomrule
      \end{tabularx}
  }  
  \vspace{0.2cm}

  \footnotesize  
  \parbox{\textwidth}{\centering
  \begin{tabular}{llllll}
  &TF:  Total Functions by keyword matching; &SF:  Average Selected Functions (per RFC); &FER: Function Extraction Rate (SF/TF);\\
  &TL:  Total Lines by keyword matching; &SL:  Average Selected Lines (per RFC); &LER: Length Extraction Rate (SL/TL)
  
  \end{tabular}
  }
\end{table*}

\begin{table*}[!htbp]
  \centering  
  \caption{Consistency Verification Experiment Results Based on GPT-4o}\label{tab:gpt4o-consistency}
  \fontsize{10pt}{12pt}\selectfont
  \resizebox{0.7\textwidth}{!}{
  \begin{tabularx}{0.9\textwidth}{c*{7}{>{\centering\arraybackslash}X}}
      \toprule
      \multirow{2}{*}{\textbf{RFC}} & \multicolumn{7}{c}{\textbf{Code Versions}} \\
      \cmidrule(lr){2-8}
        & \makecell{\textbf{Linux}\\\textbf{6.9}}  & \makecell{\textbf{Linux}\\\textbf{3.6}} & \makecell{\textbf{Linux}\\\textbf{2.6.39}} & \makecell{\textbf{Android}\\\textbf{4.19}} & \makecell{\textbf{FreeBSD}\\\textbf{13.3}} & \makecell{\textbf{NetBSD}\\\textbf{9.4}} & \makecell{\textbf{OpenBSD}\\\textbf{7.5}}\\
      \midrule
      RFC 793  & True & True  & True  & True  & True  & True  & True  \\
      RFC 1948 & True & True  & True  & True  & True  & True  & True  \\
      RFC 6528 & \textbf{\emph{True}} & False & False & \textbf{\emph{True}}  & True  & False & False \\
      RFC 5961 & True & True  & False & True  & True  & False & True  \\
      RFC 2385 & True & True  & True  & True  & True  & True  & \textbf{\emph{False}} \\
      RFC 5925 & True & False & False & False & False & False & False \\
      RFC 1323 & True & True  & True  & True  & True  & True  & True  \\
      RFC 7323 & True & False & False & \textbf{\emph{False}} & True  & False & \textbf{\emph{True}}  \\
      \bottomrule
  \end{tabularx}
  }
  \vspace{0.2cm}

  \footnotesize  
  \parbox{0.9\textwidth}{
    “True” indicates that the model determined that the corresponding kernel code implements the respective RFC standard, whereas “False” indicates non-compliance. 
    Portions marked in bold italics highlight discrepancies between the system's judgment and the established ground truth (i.e., \textbf{\emph{False}} means the check result of our model is false, but the ground truth is true).
      }
\end{table*}

\subsubsection{Code Selection}
In this experiment, we use DeepSeek to construct knowledge graphs for each code version, which integrates all TCP-related RFC documents, along with the network-relevant portions of the kernel code.   
To reduce computational costs, we employ a caching mechanism during knowledge graph construction, analyzing the full RFC structure and content only during the initial build, and reusing the established data for subsequent graphs.   
For example, the knowledge graph construction for Linux 6.9 incurred a total cost of ¥22.28 (\$3.05), and the total cost for building graphs across all seven versions was only ¥64.18 (\$8.78), demonstrating the effectiveness of the caching strategy.   
Moreover, by adopting an incremental update approach, the system only needs to analyze the changed portions when either the code or the specification documents are updated, eliminating the need to rebuild the entire knowledge graph and thus enhancing scalability and computational efficiency in large-scale, multi-version environments.

For code extraction, a simple and naive method typically relies on simple keyword matching in file or function names, which only allows for a preliminary filtering of TCP-related code without precisely pinpointing code fragments directly associated with specific RFC sections. 
In contrast, our approach leverages a pre-constructed knowledge graph to automatically extract code segments closely related to protocol functionalities, thereby reducing irrelevant noise and providing higher-quality data for subsequent consistency analysis. 
Table ~\ref{tab:coderelextract} summarizes the key performance metrics of the code extraction process across different system versions, including the number of functions,  code lines, and extraction rates.

Although extraction performance varies due to differences in code structure, modularity, and inherent system architecture, comparisons among versions reveal that in newer Linux kernels, as code size expands, the implementation of protocol functionalities becomes more comprehensive, leading to a marked increase in both the number and rate of extracted relevant code segments. 
  
\subsubsection{Consistency Verification}

To comprehensively evaluate the effectiveness of the proposed method, we employ the GPT-4o model on the previously constructed evaluation dataset. 
As one of the state-of-the-art(SOTA) large language models, GPT-4o balances detection accuracy with resource consumption, and we use GPT-4o as the baseline model in this experiment.
The evaluation covers eight RFCs and seven different versions of kernel code. 

The results are shown in Table ~\ref{tab:gpt4o-consistency}. 
In  The experimental results show that for fundamental standards such as RFC 793, RFC 1948, and RFC 1323, all systems 
conform to the specifications;    
However, for more recent or complex standards like 
RFC 5925, 
RFC 6528, 
RFC 5961 
and RFC 7323, 
inconsistency appears among systems, even among different Linux versions.    
These findings align with the challenges the proposed method aims to address, highlighting the necessity and importance of the approach.
For instance, among Linux versions, Linux 6.9 demonstrates comprehensive support for all eight RFC standards (with partial support for RFC 6528), while older versions such as Linux 3.6 and Linux 2.6.39 exhibit notable gaps with several standards, including RFC 6528, RFC 7323, and RFC 5961 (in Linux 2.6.39).  
Moreover, in BSD series, FreeBSD 13.3 achieves the highest compliance among BSD variants, while NetBSD 9.4 and OpenBSD 7.5 exhibit incomplete support for certain standards.

\subsubsection{Summary of Vulnerabilities}
\begin{table*}[htbp]
  \centering
  \renewcommand\arraystretch{0.8}
  \caption{Summary of Identified Inconsistencies}
  \label{tab:vulnerabilities}
  \fontsize{8pt}{12pt}\selectfont
  \begin{tabularx}{1\textwidth}{
    >{\centering\arraybackslash}m{0.1\textwidth} 
    >{\centering\arraybackslash}m{0.1\textwidth} 
    >{\raggedright\arraybackslash}X
    >{\raggedright\arraybackslash}m{0.22\textwidth}
  }
  \toprule
  \textbf{System} & \textbf{RFC Number} & \textbf{Description} & \textbf{Resulted Vulnerability} \\
  \midrule
  \multirow{9}{*}{Linux 3.6} 
    & RFC 6528  & Missing secret key update mechanism in ISN generation. & TCP sequence number prediction. \\
    \cmidrule(lr){2-4}
    & RFC 5925  & Missing the support of TCP Authentication Option (TCP-AO). & Replay attack risks. \\
    \cmidrule(lr){2-4}
    & \multirow{6}{*}{RFC 7323}  & Non-RST segment timestamps are not enforced (only check \verb*|sysctl_tcp_timestamps|) and \verb*|tcp_v4_reqsk_send_ack| directly uses \verb|req->rcv_wnd| without right-shift, ignoring the window scaling factor; Random per-connection timestamp offsets are not implemented, with timestamp handling relying on \verb*|tcp_time_stamp|. & \multirow{6}{*}{\makecell[l]{ TCP sequence number prediction.\\ Data injection.}} \\
  \midrule
  \multirow{13}{*}{Linux 2.6.39} 
    & RFC 6528  & Missing secret key update mechanism in ISN generation.  & TCP sequence number prediction. \\
    \cmidrule(lr){2-4}
    & \multirow{4.8}{*}{RFC 5961}  & Missing Challenge ACK mechanism for invalid RST/SYN segments; SYN segments in synchronized states trigger RST instead of a challenge ACK; 
    Missing RST/SYN sequence exact sequence number matching validation;
    Missing ACK throttling mechanism. & \multirow{4.4}{*}{\makecell[l]{RST spoofing attack.\\Blind in-window attack.\\ACK injection attack.}} \\
    \cmidrule(lr){2-4}
    & RFC 5925  & Missing the support of TCP Authentication Option (TCP-AO). & Replay attack risks. \\
    \cmidrule(lr){2-4}
    & \multirow{4.8}{*}{RFC 7323}  & Non-RST segment timestamps are not enforced (only check \verb*|sysctl_tcp_timestamps|) and \verb*|tcp_v4_reqsk_send_ack| directly uses \verb|req->rcv_wnd| without right-shift; 
    Random per-connection timestamp offsets are not implemented, relying on \verb*|tcp_time_stamp|. & \multirow{4}{*}{\makecell[l]{ TCP sequence number prediction.\\ Data injection.}} \\
  \midrule
  Android 4.19
    & RFC 5925  & Missing the support of TCP Authentication Option (TCP-AO). & Replay attack risks. \\
  \midrule
  FreeBSD 13.3 
    & RFC 5925  & Missing the support of TCP Authentication Option (TCP-AO). & Replay attack risks. \\
  \midrule
  \multirow{13.5}{*}{NetBSD 9.4} 
    & RFC 6528  & Missing secret key update mechanism in ISN generation. & TCP sequence number prediction. \\
    \cmidrule(lr){2-4}
    & \multirow{4.8}{*}{RFC 5961}  & Strict sequence checks are enforced (dropping segments when \verb|th->th_seq| != \verb|tp->rcv_nxt|) and \verb|strict_order_rst| is employed to block invalid RSTs, yielding stricter RST/SYN validation but the challenge ACK window is not implemented. 
        & \multirow{4.4}{*}{\makecell[l]{RST spoofing attack.\\Blind in-window attack.\\ACK injection attack.}} \\
    \cmidrule(lr){2-4}
    & RFC 5925  & Missing the support of TCP Authentication Option (TCP-AO). & Replay attack risks. \\
    \cmidrule(lr){2-4}
    & \multirow{6}{*}{RFC 7323}  & Window retraction handling is not implemented; Privacy enhancements of per-connection timestamp offsets are not applied as raw timestamps
     are used instead; TSopt is not added to RST segments when the triggering segment included it in \verb|tcp_output| or \verb|tcp_drop|;  No explicit adjustments to \verb|snd_wnd| or \verb|rcv_wnd| are made during retraction. & \multirow{5.5}{*}{\makecell[l]{ TCP sequence number prediction.\\ Data injection.}} \\
  \midrule
  \multirow{2.5}{*}{OpenBSD 7.5} 
    & RFC 6528  & Missing secret key update mechanism in ISN generation. & TCP sequence number prediction. \\
    \cmidrule(lr){2-4}
    & RFC 5925  & Missing the support of TCP Authentication Option (TCP-AO). & Replay attack risks. \\
  \bottomrule
  \end{tabularx}
  

  \vspace{0.2cm}

  \footnotesize  
  \parbox{0.9\textwidth}{
  }
\end{table*}

As a result, we identify 15 inconsistencies 
among these systems, including ISN generation, TCP challenge acknowledgment, TCP authentication, and TCP performance extensions.
Table~\ref{tab:vulnerabilities} summarizes these findings by listing each inconsistency, the affected system version, and a brief description of the issue. 

\subsubsection{Result Analysis}
\begin{itemize}
    \item \textbf{Effectiveness of Code Selection.}
RFC 1948 and RFC 6528 incorporate a randomization mechanism for TCP Initial Sequence Numbers (ISN), with RFC 6528 serving as an enhancement over RFC 1948.
Taking Linux 3.6 as an example,  \verb|tcp_v4_init_sequence| plays a pivotal role in implementing the randomization mechanism.
We monitor and analyze its weight throughout the extraction process.
The experimental results reveal that:
\begin{itemize}
  \item Under RFC 1948, \verb|tcp_v4_init_sequence| ranked 6th;
  \item Under RFC 6528, it ascended to the top rank.
\end{itemize}

Furthermore, differential analysis based on iterative updates further corroborates the effectiveness of the proposed method:
\begin{itemize}
\item In the update from RFC 793 to RFC 1948, \verb|tcp_v4_init_sequence| ranked 3rd, following \verb|tcp_send_delayed_ack| and \verb|tcp_md5_hash_key|;
\item In the update from RFC 1948 to RFC 6528, it ascended to the top rank.
\end{itemize}

It shows that our framework can not only identify key functions corresponding to current protocol standards,
but also accurately reflect the impact of evolving standards on code structure through incremental update analysis.    

\item \textbf{Effectiveness of Differential Analysis.}
RFC 2385 extends the basic structure of TCP by introducing the TCP MD5 signature for enhanced security, and RFC 5925 further evolves the mechanism into TCP-AO to provide stronger security assurances.
Taking Linux 6.9 as an example, In our experiments, the proposed framework successfully achieved consistency assessments in the differential analysis tasks from RFC 793 to RFC 2385 and from RFC 2385 to RFC 5925, accurately capturing and verifying the evolution of functional implementation in the code.
However, when RFC 5925 was tested independently, the model returned an “Unknown” result in 4 out of 5 trials, indicating difficulties in directly determining whether the specification had been implemented in Linux 6.9.
This result is primarily attributed to two factors:
\begin{itemize}
\item \textbf{Excessive verification scope}: The modifications in TCP authentication introduced by RFC 5925 are complex, and direct detection requires covering a wide range of code, making it challenging for the model to precisely locate the corresponding implementations;
\item \textbf{Deficient semantic bridging}: The considerable semantic gap between the specification and the code implementation hinders accurate matching in direct detection.   
\end{itemize}

\item \textbf{Error Analysis.}
To understand the limitations of our framework, we conducted a post-hoc analysis of the discrepancies between our model's predictions and the ground truth.
\begin{itemize}
    \item \textbf{False Positives}: Manual inspection confirms that TCP-MD5 is indeed implemented, but the framework failed to detect it due to the compact and context-specific nature of the code, which led to low retrieval relevance. This highlights that while the framework effectively identifies prominent and semantically rich protocol logic, its ability to capture critical information from small, localized code segments remains limited.
    \item \textbf{False Negatives}: 
Taking the evaluation of RFC 6528 in Android 4.19 as a false negative example. The code contained functions with names strongly related to ISN generation (e.g., \verb|tcp_v4_init_sequence|), leading the model to a false positive. However, a detailed review of the code logic confirmed the absence of the specific secret key update mechanism mandated by RFC 6528. The framework captured the "what" (ISN generation) but missed the critical "how" (the required periodic re-keying). This indicates that while our framework performs well in establishing initial semantic mappings, its reasoning over complex, multi-step algorithmic requirements can be constrained without explicit guidance, sometimes being misled by superficial cues and overlooking deeper algorithmic inconsistencies.
\end{itemize}
\end{itemize}

\subsection{Comparative Experiments}
\subsubsection{Method Comparison}

To further evaluate the effectiveness of our approach, we design a comparative experiment of different methods, including:
\begin{itemize}
  \item \textbf{LLM}: Directly querying the LLM to determine whether a specific RFC standard is implemented, relying solely on the model’s internal knowledge.
  \item \textbf{Naïve LLM}: Providing both the RFC document and the full source code to the LLM, allowing it to analyze and determine compliance.
  \item \textbf{LLM+CS}: Utilizing code selection (CS) to extract relevant code segments for comparison with RFC requirements.
  \item \textbf{LLM+DA}: Employing differential analysis (DA) by comparing incremental updates in RFCs with code implementations.
  \item \textbf{Ours}: Combining both code selection and differential analysis.
\end{itemize}

The results for each method is shown in Table ~\ref{tab:method_comparison}.

\begin{table}[!htbp]
  \centering  
  \caption{Comparison of Consistency Detection Methods}
  \label{tab:method_comparison}
  \fontsize{12pt}{16pt}\selectfont
  \resizebox{3.35in}{!}{
      \begin{tabular}{lccccc}
      \toprule
      \textbf{Metrics} & \textbf{LLM} & \textbf{Na\"ive LLM} & \textbf{LLM+CS} & \textbf{LLM+DA} & \textbf{Ours} \\
      \midrule
      Accuracy & 64.3\% & 48.2\% & 55.4\% & 64.3\% & 71.4\% \\
      Recall & 44.4\% & 94.4\% & 88.9\% & 100\% & 94.4\% \\
      Precision & 47.0\% & 37.8\% & 41.0\% & 47.4\% & 53.2\% \\
      F1 Score & 0.455 & 0.540 & 0.561 & 0.643 & 0.680 \\
      Total Token & / & 28.8 M & 16.3 M & 32.1 M & 19.3 M \\
      Total Cost & / & \$4.41 & \$2.53 & \$5.03 & \$3.06 \\
      Total Time & / & 3h 22m & 2h 13m & 4h 15m & 3h 39m \\
      \bottomrule
      \end{tabular}%
  }
\end{table}

The LLM-only method, which relies solely on the model’s internal knowledge, demonstrates limited effectiveness.            
This is primarily due to the model’s inability to fully recall details of different system versions, leading to guesswork and misjudgments in many cases.
The Naïve LLM approach, which directly feeds the entire source code into the model, achieves a high recall rate (94.4\%).             
However, since excessive irrelevant information is introduced, the method suffers from a significantly low precision (37.8\%), resulting in a suboptimal F1 score.

In contrast, the LLM+CS method effectively removes unrelated code and improves accuracy to 55.4\%, with an F1-score of 0.561, demonstrating the benefits of filtering techniques in consistency verification.
On the other hand, the LLM+DA method further enhances recall to 100\% while also improving accuracy to 64.3\%, yielding an F1-score of 0.643.
This suggests that differential analysis can complement direct matching approaches and enhance the comprehensiveness of detection.


Ultimately, our framework integrates code selection and differential analysis, achieving the best overall performance. It improves accuracy to 71.4\% (an absolute improvement of 23.2\% over the vanilla LLM baseline of 48.2\%) and attains an F1-score of 0.680 (an absolute gain of 0.140 over the baseline F1 of 0.540)

In terms of computational overhead, the Naïve LLM method incurs a high token consumption of 28.8M.             
The LLM+CS method is the most efficient in terms of resource consumption, reducing token usage to 16.3M, but its lower recall rate limits overall effectiveness.      
The LLM+DA method, while achieving the highest recall, further increases token consumption to 32.1M due to additional differential comparisons, leading to the highest computational cost.     
In contrast, our method balances accuracy and computational efficiency, reaching the relatively best results with token consumption of only 19.3M and reducing overall computational cost and time cost.



\subsubsection{LLM Comparison}
To assess the impact of different LLMs on consistency analysis results, we conduct a comparative experiment evaluating multiple models. 
This experiment compares five models: GPT-4o, GPT-4o-mini, o3-mini~\cite{o3_mini}, DeepSeek-V3, and DeepSeek-R1~\cite{deepseek_r1}, analyzing their influence on the differential analysis.
The results for each LLM are shown in Table ~\ref{tab:model_comparison}.
\begin{table}[!htbp]
  \centering  
  \caption{Comparison of Different LLMs for Consistency Verification}
  \label{tab:model_comparison}
  \fontsize{12pt}{16pt}\selectfont
  \resizebox{3.35in}{!}{
      \begin{tabular}{lccccc}
      \toprule
      \textbf{Metrics} & \textbf{GPT-4o} & \textbf{GPT-4o-mini} & \textbf{o3-mini} & \textbf{DeepSeek-V3} & \textbf{DeepSeek-R1} \\
      \midrule
      Accuracy & 91.1\% & 71.4\% & 82.1\% & 87.5\% & 94.6\% \\
      Recall & 83.3\% & 94.4\% & 94.4\% & 94.4\% & 94.4\% \\
      Precision & 88.2\% & 53.2\% & 65.4\% & 73.9\% & 89.5\% \\
      F1 Score & 0.857 & 0.680 & 0.773 & 0.829 & 0.919 \\
      Total Token & 19.2 M & 19.3 M & 19.9 M & 11.4 M & 12.1 M \\
      Total Cost & \$49.6 & \$3.06 & \$22.5 & \$2.07 & \$4.28 \\
      Total Time & 5h 56m & 3h 39m & 4h 2m & 8h 32m & 10h 30m \\
      \bottomrule
      \end{tabular}%
  }
\end{table}

This experiment evaluates the performance of various LLMs, focusing on detection accuracy and computational cost.  
The results show a clear trade-off between model performance and efficiency.  
Generally, within the same model series, smaller models tend to be more cost-efficient, while larger models deliver higher accuracy. 

Based on the experimental results, we conduct a comparative analysis of the models from the following four dimensions:
\begin{enumerate}
  \item \textbf{Comparison of SOTA Models} \\
  GPT-4o, DeepSeek-V3, and DeepSeek-R1 are all SOTA LLMs, but they differ in architecture and inference methods. Among them, DeepSeek-R1 achieves the best performance in the detection task, with an accuracy of 94.6\% and an F1 score of 0.919, surpassing GPT-4o and DeepSeek-V3. This result suggests that inference-optimized models can achieve higher consistency detection accuracy. While DeepSeek-V3 has slightly lower accuracy and F1 score, its computational cost is the lowest (\$2.07), making it a competitive choice in low-cost environments.

  \item \textbf{Comparison of Same-Series Models} \\
  Both GPT-4o and GPT-4o-mini belong to the same series, but GPT-4o significantly outperforms GPT-4o-mini, with an accuracy advantage of 19.7\% and an F1 score increase of 0.177. This result indicates that reducing model size has a substantial impact on consistency detection accuracy. However, GPT-4o-mini’s computational cost is only 6.2\% of GPT-4o (\$3.06 vs. \$49.6), making it a viable option for resource-constrained scenarios.

  \item \textbf{Comparison of Small Models} \\
  Both GPT-4o-mini and o3-mini are categorized as small models, but o3-mini significantly outperforms GPT-4o-mini, with 10.7\% higher accuracy and 12.2\% improvement in precision. This experiment suggests that even among small models, reasoning model can notably enhance verification performance. A similar trend is observed when comparing DeepSeek-V3 and DeepSeek-R1.

  \item \textbf{Comparison of Reasoning Models} \\
  Both o3-mini and DeepSeek-R1 are reasoning models, but as a larger-scale reasoning model, DeepSeek-R1 demonstrates significantly superior performance, achieving 12.5\% higher accuracy, 24.1\% higher precision and 0.146 higher F1 score. This indicates that larger reasoning models can enhance verification performance.
\end{enumerate}

In summary, the experimental results demonstrate that DeepSeek-R1 emerges as the best performer, followed by GPT-4o and DeepSeek-V3, while smaller models like GPT-4o-mini and o3-mini offer cost-effective alternatives with reasonable accuracy.   
These findings guide model selection, allowing users to optimize accuracy, computational cost, and inference speed based on specific application needs.

\subsection{Empirical Study}

In this experiment, we compare the ISN–generation algorithms of two recent kernel releases, Linux 6.9 and FreeBSD 13.3. According to RFC 6528, it should be

\begin{equation}
  \begin{split}
    ISN = &\, M + F(\text{localIP}, \text{localPort}, \\
                  &\, \text{remoteIP}, \text{remotePort}, \text{secretKey})
  \end{split}
  \end{equation}
where $M$ is a monotonically increasing counter and $F$ generates a hash value by the secret key and socket information. 
Both kernels implement this core requirement. 
However, as RFC 6528 notes, the security of $F$ is limited by the entropy of the secret key itself: an adversary who observes sufficiently many ISNs can mount a brute–force search against a fixed key. 
Consequently, the RFC recommends periodic or conditional re-seeding of the secret key to narrow its exposure window.

\begin{figure}[!t]
  \centering
  \includegraphics[width=3.49in]{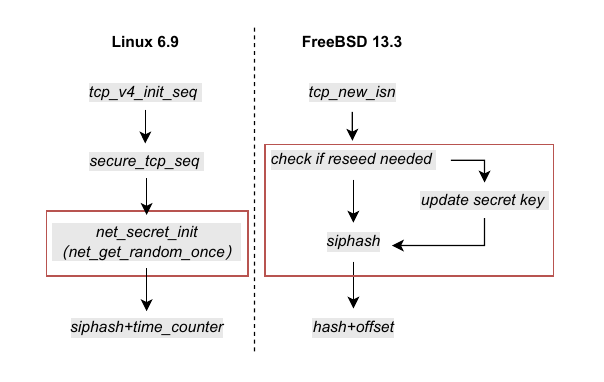}
  \caption{ISN generation algorithm between Linux 6.9 and FreeBSD 13.3} 
  \label{fig:isn_generation} 
  \end{figure}

As shown in Figure ~\ref{fig:isn_generation}, our study shows that Linux 6.9 invokes \verb|net_get_random_once|
only once at system startup (within \verb|net_secret_init|),
and thereafter leaves the secret key unchanged for the kernel’s lifetime. 
By contrast, FreeBSD 13.3 embeds a “check if reseed needed” step in \verb|tcp_new_isn|, whenever the cumulative number of ISNs or elapsed time exceeds a configurable threshold, the kernel fetches fresh entropy from a secure source to regenerate its secret key before each subsequent SipHash computation. 

These findings demonstrate that—even in their latest versions—widely deployed kernels may still diverge from RFC standards.
Such inconsistencies are not merely theoretical concerns but can introduce exploitable vulnerabilities in real-world systems.
When the secret key used in ISN generation remains static over long system uptimes, adversaries may conduct brute-force or inference attacks by observing ISNs over time, as outlined in RFC 6528.
Historically, a series of attacks have been enabled by such implementation inconsistencies. Cao et al.~\cite{caoyue2016} demonstrated that the lack of consistency between the requirements of RFC 5961 for Challenge ACKs and the implementation in Linux resulted in severe TCP hijacking attacks.
Feng et al.~\cite{feng2020} found that related inconsistencies on mixed IPID assignment in IP/TCP stack implementations enabled novel off-path TCP exploits. Yang et al.~\cite{yang2024} and Feng et al.~\cite{feng2025} exploited inconsistencies in the sequence number validation methods for TCP RST packets between intermediate routers and end hosts, which enabled local TCP hijacking or remote TCP DoS attacks in NAT-enabled networks.
These combined findings underscore that the lack of systematic consistency checks between protocol specifications and their implementations can leave systems vulnerable to subtle yet practical attacks. 

However, most previous discoveries of such inconsistencies have relied heavily on expert domain knowledge and manual analysis, which is time-consuming, difficult to scale across multiple versions and platforms, and prone to omissions.
To address this, our work significantly enhances the automation and scalability of detecting such inconsistencies. By leveraging large language models (LLMs), knowledge graph–driven protocol mapping, and incremental differential analysis, our framework enables precise and efficient detection of RFC–implementation mismatches across multiple OS versions and protocol layers. This approach not only reduces the reliance on manual auditing but also allows for proactive identification of latent vulnerabilities, thereby improving the security assurance of TCP/IP stacks in practice.

\noindent\textbf{Ethical Considerations.}
Our framework performs static analysis solely on open-source kernel code within isolated local environments, without capturing real network traffic or user data. 
It is intended strictly for academic research and teaching and may only be executed with explicit administrative permission in controlled settings. 
To prevent misuse, all analysis is logged and restricted to authorized environments. 
Any vulnerabilities discovered will follow a responsible-disclosure process: issues are reported privately to the relevant maintainers, and exploit details are withheld until patches are released. 
Users are expected to comply with applicable laws and institutional policies when using this tool.

\section{Conclusion}
\subsection{Summary of Contributions}

The rapid evolution of protocol specifications poses significant challenges for maintaining code consistency, exacerbated by unstructured documentation and complex implementations.  
Traditional verification methods which rely on expert knowledge and exhaustive full-scale analysis, turned out to be inefficient and unscalable in dynamic, large-scale network environments.

To address these challenges, we propose a novel consistency verification method of protocol specification, which integrates text parsing, knowledge graphs, and differential analysis. 
This approach establishes a comprehensive framework for parsing protocol specifications and analyzing their consistency with code implementations. 
These contributions enable practical, accurate, and scalable protocol consistency verification for real-world systems.

\begin{itemize}
\item Addressing RQ1 (Specification-Code Correlation): By constructing a knowledge graph that aligns protocol states, events, and boundary conditions with code functions, we establish bidirectional mappings between RFC text chunks and code segments.

\item Addressing RQ2 (Efficiency at Scalability): By focusing on RFC update chains and their impacted code regions, the incremental verification mechanism reduces computational overhead and makes it suitable for large-scale verification.

\item Addressing RQ3 (Semantic Gap Bridging): By constructing a structured triplet dataset and leveraging it to guide LLMs in generating protocol-aware intermediate representations, our framework serves as a unified abstraction layer that significantly bridges the specification-code semantic gap and enhances verification accuracy.
\end{itemize}


\subsection{Limitations and Future Work}

Although the proposed framework has made progress in verifying the consistency between protocol implementations and their specifications, several areas still require improvement in real-world applications.         Future work will focus on the following aspects:

\begin{enumerate}
  \item \textbf{Multi-Protocol Adaptability}

The current framework primarily analyzes a single protocol within the TCP/IP stack.         
In real-world scenarios, interactions among different protocols are far more complex and may hide additional vulnerabilities.         Future work will develop a pluggable protocol adaptation layer to quickly integrate new protocol specifications and explore cross-layer semantic correlation methods to capture security risks in inter-protocol interactions.

\item \textbf{Optimization of the Differential Analysis Model}

The current differential analysis approach mainly relies on LLMs and knowledge graphs, which may not fully meet the requirements for large-scale consistency verification.
Future research will seek to integrate static analysis, dynamic analysis, and symbolic execution techniques to refine the identification of various inconsistencies (such as missing functionality and erroneous implementations), and to optimize log analysis modules for more precise detection, thereby enhancing overall accuracy and efficiency.

\item \textbf{Enhancement of Automatic Repair Capabilities}

The present system focuses on detecting discrepancies between protocol implementations and specifications without providing automated repair functions.
Future work will explore program synthesis and automated code completion techniques to generate RFC-compliant code suggestions, and design an automated validation pipeline—integrating fuzz testing and model checking—to ensure the safety and effectiveness of generated patches, ultimately reducing maintenance costs and enhancing system security.
\end{enumerate}

\bibliographystyle{IEEEtran}
\bibliography{reference}
\end{document}